\documentclass{IEEEtran}
\usepackage{amsmath,amsfonts}
\usepackage{color}
\usepackage{array}
\usepackage{algorithmic}
\usepackage[linesnumbered,ruled,vlined]{algorithm2e}
\usepackage[caption=false,font=normalsize,labelfont=sf,textfont=sf]{subfig}
\usepackage{textcomp}
\usepackage{stfloats}
\usepackage{url}
\usepackage{times}
\usepackage{verbatim}
\usepackage{graphicx}
\usepackage[nocompress]{cite}
\usepackage{longtable}
\usepackage[utf8]{inputenc}
\usepackage[T1]{fontenc}
\usepackage{lipsum}
\usepackage{multirow}
\usepackage{mathtools}
\usepackage{titlesec}
\usepackage{hyperref}

\usepackage[utf8]{inputenc}
\usepackage[linesnumbered,ruled,vlined]{algorithm2e}
\usepackage{amsmath, amssymb}
\makeatletter
\renewcommand\paragraph{\@startsection{paragraph}{4}{\z@}%
                                    {1.3ex plus 1.5ex minus 0.2ex}%
                                    {0.7ex plus .2ex}%
                                    {\normalfont\normalsize\itshape}}
\renewcommand\subparagraph{\@startsection{subparagraph}{5}{\z@}%
                                       {1.3ex plus 1.5ex minus 0.2ex}%
                                       {0.7ex plus .2ex}%
                                       {\normalfont\normalsize\itshape}}
\makeatother

\begin{document}

\linespread{0.999}
\title{Study of Robust Power Allocation for User-Centric Cell-Free
Massive MIMO Networks}

\author{Saeed Mashdour$^{\star}$, Saeed Mohammadzadeh$^{\dagger}$, André R. Flores $^{\star}$, Shirin Salehi $^{\star\star}$, Rodrigo C. de Lamare $^{\star,\dagger}$, Anke Schmeink $^{\star\star}$ \vspace{-1.05em} 

\thanks{Saeed Mashdour, André R. Flores and  Rodrigo C. de Lamare are with Pontifical Catholic University of Rio de Janeiro, 22541-041 Rio de Janeiro, Brazil, Rodrigo C. de Lamare and Saeed Mohammadzadeh are with the University of York, UK, Shirin Salehi and Anke Schmeink are with Chair of Information Theory and Data Analytics, RWTH Aachen University, 52062 Aachen, Germany. Emails are smashdour@gmail.com, andre.flores@ucsp.edu.pe, saeed.mohammadzadeh@york.ac.uk, delamare@puc-rio.br, \{shirin.salehi, anke.schmeink\}@inda.rwth-aachen.de. This work was supported by CNPQ, FAPESP and FAPERJ. }}

\maketitle
\thispagestyle{empty}
\begin{abstract}
In cell‐free massive multiple-input multiple-output (MIMO) networks, robust resource allocation is critical to ensure reliable system performance in the presence of channel uncertainties resulting from imperfect channel state information (CSI). In this work, we propose a robust power allocation method that formulates the power optimization problem into a least‐squares framework, enhanced by Tikhonov regularization to mitigate the adverse effects of channel estimation errors. We integrate our approach with zero-forcing precoding, enabling a design that is both computationally efficient and resilient to CSI imperfections. Numerical results indicate that the proposed method outperforms existing non-robust techniques while benefiting from low computational overhead, making it well-suited for large-scale deployments under CSI uncertainty.
\end{abstract}

\section{Introduction} 

Cell-free massive multiple-input multiple-output (MIMO) (CF-mMIMO) networks are envisioned as a fundamental enabler of future wireless systems due to their potential for uniformly high service quality through spatial diversity and interference mitigation \cite{ngo2015cell, interdonato2019ubiquitous} and further enhance co-located massive multiple-input multiple-output (mMIMO) systems \cite{mmimo,wence}. In these networks, a large number of distributed access points (APs) jointly serve users without relying on cell boundaries, enabling gains in spectral and energy efficiency \cite{bjornson2020scalable}.

Effective allocation of resources in CF-mMIMO, particularly power \cite{mohammadzadeh2025association,11311474,jpba,tds,tds2}, is crucial to maximizing system performance \cite{nayebi2017precoding, van2016joint}. However, a major performance bottleneck in CF-mMIMO is the imperfect knowledge of channel state information (CSI) due to estimation errors, pilot contamination, and feedback delays. These imperfections degrade the system performance by introducing mismatches in transmit processing and resource allocation strategies \cite{interdonato2019ubiquitous, flores2022robust}. Therefore, robust strategies that can deal with channel uncertainty and mitigate its effects are essential \cite{locsme,okspme,lrcc,rdrls,rthp,palhares2021robust,rsbd,rsthp,rrs}.

This paper investigates a robust power allocation strategy based on Tikhonov regularization, which is widely used to stabilize ill-posed inverse problems and reduce the sensitivity to noise and perturbations \cite{hofmann2013regularization, zou2005regularization}. Specifically, we consider a least-squares (LS) formulation that takes advantage of an equivalent linear model of the received signal, excluding the estimation error during optimization. Robustness is then introduced through a regularization term that penalizes power levels, which are prone to amplify channel errors.

%\textcolor{blue}{Saeed, is it essential to use a multiuser scheduling algorithm? You can simply use a chosen number of users for precoding and power allocation. Woult it not be simpler?}

{We first schedule a subset of users using the clustered-enhanced subset greedy (C-ESG) algorithm \cite{mashdour2022enhanced}, such that in each coherence block, C-ESG selects $n$ user equipments (UEs) out of the $K$ candidate UEs via a greedy procedure based on the channel power, and the same scheduled user set is used for all schemes.} Given the user set, we apply the proposed power allocation strategy. This sequential approach enables power allocation to focus only on the active users, thereby simplifying the optimization process and avoiding computationally intensive optimization methods under channel uncertainty. Compared to former robust approaches, such as those proposed in \cite{mashdour2024robust, mashdour2025robust}, the proposed method offers comparable computational complexity while maintaining robustness to channel estimation errors. Its closed-form nature makes it an effective and practical alternative for robust design.

The remainder of this paper is organized as follows. Section~\ref{SYS.mod} presents the system model and formulates the optimization problem. In Section~\ref{Solut.}, we propose a least-squares-based method to solve the power allocation problem under channel uncertainty. In Section~\ref{Analys}, we derive the sum-rate expression for the system and analyze the computational complexity. Section~\ref{Simul.} presents and discusses the numerical results. Finally, Section~\ref{Conclud.} concludes the paper.

Notation: $\mathbf{I}_n$ denotes the $n \times n$ identity matrix. The complex normal distribution is denoted by $\mathcal{CN}(\cdot,\cdot)$. Superscripts $^{T}$, $^{*}$, and $^{H}$ represent transpose, complex conjugate, and Hermitian transpose, respectively. The space of complex $M \times N$ matrices is $\mathbb{C}^{M \times N}$. $\text{Tr}(\cdot)$ is the matrix trace. $\|\cdot \|_F$ denotes the Frobenius norm. $\text{diag}(\mathbf{x})$ forms a diagonal matrix from vector $\mathbf{x}$, and $\text{diag}(\mathbf{X})$ extracts the diagonal of matrix $\mathbf{X}$. The expectation operator is $\mathbb{E}[\cdot]$, and $[\mathbf{X}]_{m,n}$ denotes the $(m,n)$-th entry of $\mathbf{X}$.

\section{System Model and Problem Formulation} \label{SYS.mod}

\subsection{System Model}

We consider the downlink of a CF-mMIMO network consisting of $L$ distributed APs, each equipped with $N$ antenna elements, jointly serving $K$ single-antenna UEs. The system operates under a user-centric cell-free (UCCF) approach, where each UE is dynamically associated with a subset of APs based on large-scale fading (LSF) conditions \cite{decidd,iddllr,iddocl}. The total number of UE is assumed to be significantly larger than the total number of available AP antennas, i.e., $K \gg M = LN$. To maintain reliable system operation, only a subset of $n \leq M$ UEs is scheduled for transmission in each resource block by using the C-ESG multiuser scheduling technique. The channel matrix between the APs and the $n$ scheduled UEs is represented as $\mathbf{G} = \hat{\mathbf{G}} + \tilde{\mathbf{G}} \in \mathbb{C}^{LN \times n}$, where $\hat{\mathbf{G}}$ is the channel estimate and $\tilde{\mathbf{G}}$ models the channel estimation error. The error $\tilde{\mathbf{G}}$ is assumed to be a complex Gaussian random matrix with zero mean and known variance, capturing channel uncertainty~\cite{nayebi2017precoding}. The element $g_{mk} = [\mathbf{G}]_{m,k}$ denotes the channel coefficient between the $m$th AP antenna and the $k$th UE and is given by 
\begin{equation} \label{eq.gI}
\begin{split}
    g_{{mk}} = & \hat{g}_{mk}+\tilde{g}_{mk} \\
    = & \sqrt{1-\alpha}\sqrt{\beta_{mk}}h_{mk} + \sqrt{\alpha}\sqrt{\beta_{mk}}\tilde{h}_{mk},
\end{split}
\end{equation}
where $\alpha$ is the CSI imperfection factor $(0 < \alpha < 1)$, $\beta_{mk}$ is the known LSF coefficient between AP antenna $m$ and UE $k$, and $h_{mk}, \tilde{h}_{mk} \sim \mathcal{CN}(0,1)$ are the small-scale fading and estimation error terms, respectively. These are assumed to be i.i.d. and mutually independent~\cite{flores2023clustered, nayebi2017precoding, mishra2022rate}.

{In the UCCF framework, AP selection follows the LSF-based clustering in \cite{ammar2021downlink}, where the LSF is modeled as $\beta_{mk}=\mathrm{PL}_{mk}10^{\sigma_{\mathrm{sh}}z_{mk}/10}$ with $\sigma_{\mathrm{sh}}=8$~dB, $z_{mk}\sim\mathcal{N}(0,1)$ and $\mathrm{PL}_{mk}$ generated from the distance-based pathloss model in~\cite{ngo2017cell,tang2001mobile}. For each UE $k$, the serving AP set is $U_k = \{m:\beta_{mk}\ge\lambda_{\mathrm{lsf}}\}\cup\{\arg\max_m \beta_{mk}\}$, where $\lambda_{\mathrm{lsf}}$ is the average of all network LSF coefficients, so that each UE is associated with a cluster of APs whose $\beta_{mk}$ exceeds this threshold and the AP with the strongest LSF is always included to guarantee minimum service coverage.} This results in a sparse channel matrix $\mathbf{G}_a$ where only the elements corresponding to active AP-UE pairs in each column are non-zero. To enhance signal transmission, linear precoding is applied at the APs using a zero-forcing (ZF) precoding strategy. The corresponding precoding matrix is denoted by $\mathbf{P}_a \in \mathbb{C}^{LN \times n}$, which is a function of the sparse estimated channel matrix. The downlink received signal vector $\mathbf{y}_a \in \mathbb{C}^{n \times 1}$ at the scheduled UEs is expressed as \vspace{-2mm}
\begin{equation}\label{UCCF.sig}
\begin{split}
    \mathbf{y}_a
    &= \sqrt{\rho_f}\,\mathbf{G}^{{T}}\mathbf{P}_a\,\mathbf{x} + \mathbf{w}\\[2mm]
    &= \sqrt{\rho_f}\,\hat{\mathbf{G}}^{{T}}\mathbf{P}_a\,\mathbf{x}
       \;+\; \sqrt{\rho_f}\,\tilde{\mathbf{G}}^{{T}}\mathbf{P}_a\,\mathbf{x}
       \;+\; \mathbf{w},
\end{split}
\end{equation}
where $\rho_f$ is the downlink transmit power, $\mathbf{x}$ is the $n \times 1$ transmit symbol vector with normalized power, $\mathbf{w} \sim \mathcal{CN}(\mathbf{0}, \sigma_w^2 \mathbf{I}_n)$ is the additive white Gaussian noise vector, $\mathbf{P}_a$ denotes the zero‑forcing precoding matrix. In \eqref{UCCF.sig}, we can use $\mathbf{G}_a^{{T}}\mathbf{P}_a$ as well, because it implicitly selects the same channel entries as $\mathbf{G}^{{T}}\mathbf{P}_a$, however, since the physical propagation channel is $\mathbf{G}$, the full channel is used instead of its sparsified version $\mathbf{G}_a$. Assuming Gaussian signaling and statistical independence between $\mathbf{x}$ and $\mathbf{w}$, the upper bound on the achievable sum-rate (\text{SR}) under imperfect channel knowledge for the scheduled UEs is expressed as follows \cite{mashdour2025robust}:\vspace{-2mm} %(derivation is given in Section \ref{sum-rate}) 
\begin{equation}\label{eq:RCF}
    \text{SR} = \log_2\left(\det\left[\mathbf{R}_{UC_{\tilde{\mathbf{G}}}} + \mathbf{I}_K\right]\right),
\end{equation}
where $\mathbf{I}_K$ is the identity matrix, $\mathbf{R}_{UC_{\tilde{\mathbf{G}}}} = \rho_f \hat{\mathbf{G}}^T \mathbf{P}_a \mathbf{P}_a^H \hat{\mathbf{G}}^* \left( \mathbf{R}_{\tilde{\mathbf{G}}} \right)^{-1}$, and \vspace{-2mm}
\begin{equation}
    \mathbf{R}_{\tilde{\mathbf{G}}} = \mathbb{E}_{\tilde{\mathbf{G}}} \left[ \rho_f \tilde{\mathbf{G}}^T \mathbf{P}_a \mathbf{P}_a^H \tilde{\mathbf{G}}^* \right] + \sigma_w^2 \mathbf{I}_K.
\end{equation}

\vspace{-4mm}
\subsection{Problem Formulation}
\vspace{-1mm}
After scheduling the intended users, we proceed with power allocation to optimize the system performance. The received signal at the \( n \) scheduled users is then expressed as follows: 
\begin{equation}\label{eq:signal}
\begin{split}
\mathbf{y}_a &=  \sqrt{\rho_f}\hat{\mathbf{G}}^T\,\mathbf{P}_a \mathbf{x} + \sqrt{\rho_f}\tilde{\mathbf{G}}^T\,\mathbf{P}_a \mathbf{x} + \mathbf{w}\\
& =\sqrt{\rho_f}\hat{\mathbf{G}}^T\,\mathbf{W}\mathbf{D} \mathbf{x} + \sqrt{\rho_f}\tilde{\mathbf{G}}^T\,\mathbf{P}_a \mathbf{x} + \mathbf{w},
\end{split}
\end{equation}
where $\mathbf{P}_a = \mathbf{W}\mathbf{D}$, with $\mathbf{W}\in\mathbb{C}^{NL \times {n}}$ as the normalized precoding matrix and $\mathbf{D}=\textup{diag}\left ( \mathbf{d} \right )$ as the diagonal power allocation matrix with the power allocation factors given by $\mathbf{d}=\left [ {d_{1}} \ {d_{2}} \ \cdots \  {d_{{n}}} \right ]^T$. Accordingly, the error between the transmitted signal and the received signal is defined as 

\begin{equation}\label{eq:error}
\varepsilon = \|\mathbf{x} - \mathbf{y}_a\|^2.
\end{equation}
Since channel estimation errors arise due to CSI imperfections, we aim to design a power allocation strategy that remains robust against these errors. To achieve this, we formulate the following optimization problem, which employs a worst-case robust power allocation technique: \vspace{-2mm}
\begin{equation}\label{eq:robust_problem}
\begin{aligned}
& \min_{\mathbf{d}} \; \max_{\tilde{\mathbf{G}} } \; \mathbb{E}\left[\varepsilon\right] \\
& \text{subject to } \|\mathbf{W}\,\operatorname{diag}(\mathbf{d})\|_F^2 \le P, \\
&\quad \quad \quad \quad
\mathbf{d} \succeq 0,
\\
&\quad \quad \quad \quad
\kappa_1\leq \| \tilde{\mathbf{G}}\|^{2}\leq \kappa_2.
\end{aligned}
\end{equation}
where $P$ is the upper limit of the signal covariance matrix $\text{Tr}\left ( \mathbf{C}_{\mathbf{x}}\right )\leq P$, and $\kappa_1$, $\kappa_2$ bound the channel estimation error matrix norm.
{We choose $\kappa_1$ and $\kappa_2$ so that the level of CSI imperfection $\alpha$ stays within a moderate range, neither close to zero nor too large, {where $\kappa_1>0$ excludes unrealistically small (near error-free) estimation errors and $\kappa_2$ prevents unrealistically large errors for which the channel estimate would be unusable.} This ensures that the worst-case formulation in \eqref{eq:robust_problem} remains conservative enough to cope with CSI uncertainties while maintaining a balanced and realistic configuration.
}

\section{Proposed least-squares Based Solution} \label{Solut.}

\subsection{Robust Power Allocation Vector Derivation}

In our robust power allocation problem, we aim to minimize the expected squared error between the transmitted signal $\mathbf{x}$ and the received signal $\mathbf{y}_a(\mathbf{d}, \tilde{\mathbf{G}})$ as
\begin{equation}
  \min_{\mathbf{d}} \; \max_{\tilde{\mathbf{G}}} \; \mathbb{E}\left[\|\mathbf{x}-\mathbf{y}_a(\mathbf{d}, \tilde{\mathbf{G}})\|^2\right].  
\end{equation}
Directly solving this problem is difficult because the error term $\tilde{\mathbf{G}}$ is random, and its statistics (although known) make the problem stochastic.

To simplify the approach, we first set aside the effect of the CSI error and consider only the estimated channel. In other words, we use the signal defined by
\vspace{-2mm}
\begin{equation} \label{y_G_hat}
    \mathbf{y}_a(\mathbf{d}) 
= \sqrt{\rho_f} \hat{\mathbf{G}}^T \mathbf{W} \,\mathrm{diag}(\mathbf{d}) \mathbf{x} + \mathbf{w},
\end{equation}
and perform a power allocation per transmitted symbol, implementing a least-squares (LS) approach. 
We exploit a fundamental property of diagonal matrices, which states that\vspace{-2mm}
 \begin{equation}
     \mathrm{diag}(\mathbf{d}) \mathbf{x} = \mathrm{diag}(\mathbf{x}) \mathbf{d},
 \end{equation}
and rewrite (\ref{y_G_hat}) as\vspace{-2mm}
\begin{equation} \label{y_a_diag_x}
    \mathbf{y}_a(\mathbf{d}) 
= \sqrt{\rho_f} \hat{\mathbf{G}}^T \mathbf{W} \,\mathrm{diag}(\mathbf{x}) \mathbf{d} + \mathbf{w}.
\end{equation}
Thus, defining \vspace{-2mm}
\begin{equation}
\mathbf{A} = \sqrt{\rho_f} \hat{\mathbf{G}}^T \mathbf{W} \,\mathrm{diag}(\mathbf{x}),
\end{equation}
we seek to minimize the following \vspace{-2mm}
\begin{equation}
    \begin{split}
        \mathbb{E}\left[\bigl\|\mathbf{x} - \mathbf{y}_a(\mathbf{d})\bigr\|^2\right]
&= \mathbb{E}\left[\bigl\|\mathbf{x} - (\mathbf{A}\mathbf{d} + \mathbf{w})\bigr\|^2\right] \notag\\
&= \mathbb{E}\Bigl[\bigl(\mathbf{x} - \mathbf{A}\mathbf{d} - \mathbf{w}\bigr)^{H}\bigl(\mathbf{x} - \mathbf{A}\mathbf{d} - \mathbf{w}\bigr)\Bigr]
    \end{split}
\end{equation}
Let \(\mathbf{r} = \mathbf{x} - \mathbf{A}\mathbf{d}\), taking expectations gives \vspace{-2mm}
\begin{equation}
    \begin{split}
        & \mathbb{E}\Bigl[\bigl(\mathbf{x} - \mathbf{A}\mathbf{d} - \mathbf{w}\bigr)^{H}\bigl(\mathbf{x} - \mathbf{A}\mathbf{d} - \mathbf{w}\bigr)\Bigr]=
\\
&\mathbb{E}\left[(\mathbf{r}-\mathbf{w})^{H}(\mathbf{r}-\mathbf{w})\right]=\\
& \mathbb{E}\bigl[\mathbf{r}^{H}\mathbf{r}\bigr]
-\mathbb{E}\bigl[\mathbf{r}^{H}\mathbf{w}\bigr]
-\mathbb{E}\bigl[\mathbf{w}^{H}\mathbf{r}\bigr]
+\mathbb{E}\bigl[\mathbf{w}^{H}\mathbf{w}\bigr]
    \end{split}
\end{equation}
Since $\mathbb{E}\bigl[\mathbf{w}\bigr]=\mathbf{0}$, we have $\mathbb{E}\bigl[\mathbf{r}^{H}\mathbf{w}\bigr]= 0$ and $\mathbb{E}\bigl[\mathbf{w}^{H}\mathbf{r}\bigr]=0$. In addition, since power allocation is performed per transmitted symbol and $\mathbf{x}$ is thus known at each step, we treat it as deterministic and obtain \vspace{-2mm}
\begin{equation}
    \begin{split}
       & \mathbb{E}\Bigl[\bigl(\mathbf{x} - \mathbf{A}\mathbf{d} - \mathbf{w}\bigr)^{H}\bigl(\mathbf{x} - \mathbf{A}\mathbf{d} - \mathbf{w}\bigr)\Bigr]
=\\
&\|\mathbf{x}-\mathbf{A}\mathbf{d}\|^{2} +\mathbb{E}\!\bigl[\|\mathbf{w}\|^{2}\bigr].
    \end{split}
\end{equation}
The term $\mathbb{E}\bigl[\|\mathbf{w}\|^{2}\bigr] = \sigma_w^2\,n$ is a constant that does not depend on~$\mathbf{d}$, so minimizing the expected error reduces to the following standard LS problem \vspace{-2mm}
\begin{equation}
    \min_{\mathbf{d}} \; \|\mathbf{x} - \mathbf{A}\mathbf{d}\|^{2}.
\end{equation}
To account for the impact of imperfect CSI, where channel estimation errors can degrade the solution, we incorporate a Tikhonov (ridge) regularization term. This leads to the following optimization problem: \vspace{-2mm}
\begin{equation} \label{Opt.Pr}
    \min_{\mathbf{d}} \; \|\mathbf{x} - \mathbf{A}\mathbf{d}\|^{2}+ \lambda \|\mathbf{d}\|^2.
\end{equation} 
where the parameter $\lambda$ is chosen based on the error statistics to ensure that the solution remains robust against the uncertainties in $\tilde{\mathbf{G}}$. {Motivated by the robust min–max formulation in \eqref{eq:robust_problem}–\eqref{y_G_hat}, this leads to a regularized LS problem that can be interpreted as a tractable substitute obtained by minimizing an explicit analytic upper bound on the worst-case residual under the adopted CSI uncertainty set.} Regularization is particularly beneficial when the underlying problem is ill-conditioned or when uncertainties in the system model might otherwise lead to unstable or biased solutions \cite{fuhry2012new, dou2017signal}. With channel estimation errors, the optimization problem can become sensitive to small perturbations in \(\hat{\mathbf{G}}\), leading to unfair or inefficient power distribution. The regularization term discourages the solution from having an excessively large norm, thereby smoothing the solution and making it less sensitive to these errors.

%  To address the problem in (\ref{Opt.Pr}), we first exploit a fundamental property of diagonal matrices, which states that
%  \begin{equation}
%      \mathrm{diag}(\mathbf{d}) \mathbf{x} = \mathrm{diag}(\mathbf{x}) \mathbf{d},
%  \end{equation}
% and rewrite (\ref{y_G_hat}) as
% \begin{equation} \label{y_a_diag_x}
%     \mathbf{y}_a(\mathbf{d}) 
% = \sqrt{\rho_f} \hat{\mathbf{G}}^T \mathbf{W} \,\mathrm{diag}(\mathbf{x}) \mathbf{d} + \mathbf{w}.
% \end{equation}
% Thus, defining
% \begin{equation}
% \mathbf{A} = \sqrt{\rho_f} \hat{\mathbf{G}}^T \mathbf{W} \,\mathrm{diag}(\mathbf{x}),
% \end{equation}
Considering the cost function of the problem in (\ref{Opt.Pr}) \vspace{-2mm}
\begin{equation}
J(\mathbf{d}) = \|\mathbf{x} - \mathbf{A} \mathbf{d}\|^2 + \lambda \|\mathbf{d}\|^2,
\end{equation}
and using the identity \(\|\mathbf{z}\|^2 = \mathbf{z}^H \mathbf{z}\), the cost function can be rewritten as:\vspace{-2mm}
\begin{equation}
\begin{split}
J(\mathbf{d}) 
&= (\mathbf{x} - \mathbf{A} \mathbf{d})^H (\mathbf{x} - \mathbf{A} \mathbf{d}) + \lambda\, \mathbf{d}^H \mathbf{d}  \\
&= \mathbf{x}^H \mathbf{x}
-  \mathbf{x}^H\mathbf{A}\mathbf{d} \\
&
- \mathbf{d}^H\mathbf{A}^H \mathbf{x} + \mathbf{d}^H\mathbf{A}^H \mathbf{A} \mathbf{d}+ \lambda\, \mathbf{d}^H \mathbf{d}.
\end{split}
\end{equation}
Since $J(\mathbf{d})$ is scalar, it is the same when we apply the trace operator as \vspace{-2mm}
\begin{equation}
    \begin{split}
        \operatorname{Tr}\bigl(J(\mathbf{d})\bigr) &=\operatorname{Tr}\bigl[\mathbf{x}^H \mathbf{x}\bigr]
        - \operatorname{Tr}\bigl[\mathbf{x}^H\mathbf{A}\mathbf{d}\bigr] - \operatorname{Tr}\bigl[\mathbf{d}^H\mathbf{A}^H \mathbf{x}\bigr]\\
&+\operatorname{Tr}\bigl[\mathbf{d}^H\mathbf{A}^H\mathbf{A}\mathbf{d}\bigr] + \lambda\,\operatorname{Tr}\bigl[\mathbf{d}^H\mathbf{d}\bigr]
    \end{split}
\end{equation}
To find the optimal power allocation vector $\mathbf{d}$, we take the derivative of the cost function as $\frac{\partial \textup{Tr}\left( J(\mathbf{d}) \right)}{\partial \mathbf{d}}$. The derivative of the first term $\operatorname{Tr}\bigl[\mathbf{x}^H \mathbf{x}\bigr]$ with respect to $\mathbf{d}$ is zero. For the derivative of the second term, we use $\frac{\partial}{\partial \mathbf{Z}} \operatorname{Tr}(\mathbf{B}\mathbf{Z}) = \mathbf{B}^T$ \cite{hjorungnes2007complex}, and we obtain \vspace{-2mm}
\begin{equation}
    \frac{\partial}{\partial \mathbf{d}} \operatorname{Tr}\bigl[\mathbf{x}^H\mathbf{A}\mathbf{d}\bigr] = \mathbf{A}^T\mathbf{x}^*
\end{equation}
Using $\frac{\partial}{\partial \mathbf{Z}} \operatorname{Tr}(\mathbf{Z}^H \mathbf{A}) = \mathbf{0}$, the derivative of the third term is \vspace{-2mm}
\begin{equation}
   \frac{\partial}{\partial \mathbf{d}} \operatorname{Tr}\bigl[\mathbf{d}^H\mathbf{A}^H \mathbf{x}\bigr]= \mathbf{0}
\end{equation}
For the term $\operatorname{Tr}\bigl[\mathbf{d}^H\mathbf{A}^H\mathbf{A}\mathbf{d}\bigr]$, we use the property $\frac{\partial}{\partial \mathbf{Z}}\operatorname{Tr}\left\{ \mathbf{Z} \mathbf{B}_0 \mathbf{Z}^H \mathbf{B}_1 \right\}
=\mathbf{B}_1^T \mathbf{Z}^* \mathbf{B}_0^T$ \cite{hjorungnes2007complex} and obtain 
\begin{equation}
    \frac{\partial}{\partial \mathbf{d}} \left( \mathbf{d}^H \mathbf{A}^H \mathbf{A} \mathbf{d} \right) = \mathbf{I}^T \mathbf{d}^* (\mathbf{A}^H \mathbf{A})^T = \mathbf{d} (\mathbf{A}^T \mathbf{A}^*)
\end{equation} 
because the power allocation vector $\mathbf{d}$ is real-valued. Using the same property, we obtain the following \vspace{-2mm}
\begin{equation}
    \frac{\partial}{\partial \mathbf{d}} \left( \lambda \mathbf{d}^H \mathbf{d} \right) =  \lambda \mathbf{d}^*=\lambda \mathbf{d}
\end{equation}
Thus, the derivative of the objective function with respect to the power allocation factor is obtained as follows: \vspace{-2mm}
\begin{equation}
    \frac{\partial \textup{Tr}\left( J(\mathbf{d}) \right)}{\partial \mathbf{d}} = -\mathbf{A}^T\mathbf{x}^* + \mathbf{d} \mathbf{A}^T \mathbf{A}^* + \lambda \mathbf{d}
\end{equation}
To find the optimal solution for $\mathbf{d}$, we set the gradient equal to zero $\frac{\partial}{\partial \mathbf{d}} J(\mathbf{d}) =0$. Thus, we obtain \vspace{-2mm}
\begin{equation} \label{complex_d}
\begin{split}
     &\mathbf{d} (\mathbf{A}^T \mathbf{A}^* + \lambda \mathbf{I}) = \mathbf{A}^T \mathbf{x} ^*
 \\
    &
\Rightarrow \mathbf{d} = (\mathbf{A}^T \mathbf{A}^* + \lambda \mathbf{I})^{-1} \mathbf{A}^T \mathbf{x} ^*
\end{split}
\end{equation}
Since $\mathbf{d}$ represents power allocation factors, it is constrained to be real-valued. Re-deriving the optimal solution with $\mathbf{d}\in\mathbb{R}^n$ using standard real-valued matrix calculus \cite{2008matrix} yields 
\begin{equation}
  {\mathbf{d}
  = \bigl(\mathbf{A}^H \mathbf{A} + \mathbf{A}^T \mathbf{A}^* + 2\lambda\mathbf{I}\bigr)^{-1}
    \bigl(\mathbf{A}^T \mathbf{x}^* + \mathbf{A}^H \mathbf{x}\bigr).}
\end{equation}
{Using $\mathbf{A}^H \mathbf{A} + \mathbf{A}^T \mathbf{A}^* = 2\,\mathrm{Re}\{\mathbf{A}^T \mathbf{A}^*\}$ and $\mathbf{A}^H \mathbf{x} + \mathbf{A}^T \mathbf{x}^* = 2\,\mathrm{Re}\{\mathbf{A}^T \mathbf{x}^*\}$, we obtain an expression for $\mathbf{d}$ that coincides with the real part of the solution given in \eqref{complex_d}.}

\subsection{Determination of the Regularization Parameter} 

One way to determine the optimal $\lambda$ is through the empirical method, which involves performing simulations or cross-validation to evaluate the performance of the system under different values of $\lambda$. We can generate multiple realizations of $\tilde{\mathbf{G}}$ within the specified bounds and solve the optimization problem for each realization of $\tilde{\mathbf{G}}$ for various $\lambda$ values. The performance is then computed for each case, and the value of $\lambda$ that maximizes the performance ensures a robust and efficient solution for the power allocation vector $\mathbf{d}$.

{However, for simplicity and analytical tractability, we adopt a robust LS viewpoint. Under the uncertainty set $\{\tilde{\mathbf{G}}:\kappa_1\leq \| \tilde{\mathbf{G}}\|^{2}\leq \kappa_2\}$, the estimation error induces a norm-bounded perturbation of the effective LS matrix, and a simple bound shows that the additional residual term due to this perturbation is upper-bounded (up to constants) by }% the following equation in line with robust LS interpretations of Tikhonov regularization~\cite{el1997robust}:}\vspace{-2mm}
\begin{equation} \label{lambda}
    \lambda = \rho_f\,\kappa_2\,\|\mathbf{W}\,\operatorname{diag}(\mathbf{x})\|^2,
\end{equation}
where $\kappa_2$ bounds the worst-case squared norm of the channel estimation error $\tilde{\mathbf{G}}$ {and which is in line with robust LS interpretations of Tikhonov regularization~\cite{el1997robust}}. As the error $\tilde{\mathbf{G}}$ propagates through the precoding matrix $\mathbf{W}$ and is further scaled by the signal vector (via $\textup{diag}(\mathbf{x})$ based on (\ref{y_a_diag_x})) and the transmit power $\rho_f$, its adverse impact is upper bounded by \eqref{lambda}. By choosing $\lambda$ sufficiently large, the regularization term $\lambda\|\mathbf{d}\|^2$ effectively counteracts the influence of the worst-case estimation error, ensuring that the resulting vector $\mathbf{d}$ remains robust under adverse channel conditions and adaptive to the power distribution in the system. {Consequently, the Tikhonov term $\lambda\|\mathbf{d}\|^2$ can be interpreted as a conservative upper bound on the worst-case contribution of the CSI error to the LS cost, providing a robust choice of $\lambda$ without implying global optimality to the min-max problem defined in \eqref{eq:robust_problem}.}

\subsection{{Post-Processing to Enforce Constraints}}

After computing the power allocation vector $\mathbf{d}$ from the LS formulation, we perform a projection step to ensure that the solution satisfies the system constraints. First, to enforce the total power constraint $\| \mathbf{W}\,\mathrm{diag}(\mathbf{d}) \|_F^2 \le P_{\text{max}}$, we rescale the solution if the constraint is violated as follows:\vspace{-2mm}
\begin{equation}
\mathbf{d} \leftarrow \mathbf{d} \cdot \frac{P_{\text{max}}}{\|\mathbf{W}\,\mathrm{diag}(\mathbf{d})\|_F^2}.
\end{equation}
Then, to enforce non-negativity of the power allocation {(as required by $\mathbf{d}\succeq\mathbf{0}$ in (7))}, we apply a projection of $\mathbf{d}$ onto the non-negative orthant by zeroing out any negative entries:\vspace{-2mm}
\begin{equation}
d_i \leftarrow \max(0, d_i), \quad \text{for } i = 1, \dots, n.
\end{equation}
These projection steps ensure that the final solution $\mathbf{d}$ is both power-limited and physically feasible, and they correspond to projecting the regularized LS solution onto the convex feasible set $\{\mathbf{d}:\| \mathbf{W}\,\mathrm{diag}(\mathbf{d}) \|_F^2 \le P_{\text{max}},\,\mathbf{d}\succeq\mathbf{0}\}$, which does not amplify perturbations and thus preserves the robustness provided by the regularization. Accordingly, we can summarize the proposed robust least-squares power allocation (RLSPA) solution in Algorithm \ref{alg:RLSPA}.
\vspace{-5mm}

\section{Complexity Analysis} \label{Analys}
\vspace{-2mm}  
The computational cost of RLSPA primarily comes from its matrix operations and regularized LS formulation. The construction of the matrix $\mathbf{A} = \sqrt{\rho_f} \hat{\mathbf{G}}^T \mathbf{W}\,\mathrm{diag}(\mathbf{x})$ involves a matrix multiplication with a complexity of $\mathcal{O}(Mn^2)$, where $M = LN$ denotes the total number of transmit antennas and $n$ is the number of scheduled users. The parameter $\lambda = \rho_f \kappa_2 \|\mathbf{W}\,\mathrm{diag}(\mathbf{x})\|^2$ can be computed with a matrix-vector norm operation, resulting in a cost of $\mathcal{O}(Mn)$. The most demanding step is solving the regularized LS equation $\mathbf{d} = \left( \mathbf{A}^T \mathbf{A}^* + \lambda \mathbf{I} \right)^{-1} \mathbf{A}^T \mathbf{x}^*$, which involves a matrix inversion and multiplication with a complexity of $\mathcal{O}(n^3 + Mn^2)$. Additionally, the projection step to enforce the total power constraint and non-negativity condition requires computing the Frobenius norm and adjusting the entries of $\mathbf{d}$, contributing $\mathcal{O}(Mn + n)$ to the complexity. {Since RLSPA is formulated as a symbol-level power allocation scheme, these calculations are repeated per symbol, and the overall complexity becomes $ \mathcal{O} \left ( N_{\rm sym}(n^3 + Mn^2) \right )$ where $N_{\rm sym}$ is the number of symbols per coherence interval}. In comparison, the robust gradient descent power allocation (RGDPA) method of \cite{mashdour2025robust} is iterative, each iteration resulting in a cost of $\mathcal{O}(4I_DMn^2)$ for $I_D$ iterations. The gradient descent power allocation (GDPA) method in \cite{mashdour2025robust}, which is a special case of RGDPA, has cost of $\mathcal{O}(3I_t \cdot Mn^2)$ for $I_t$ iterations. Although RGDPA has lower complexity per iteration than RLSPA, the total cost grows linearly with the number of iterations. Overall, the comparison of the RLSPA, RGDPA and GDPA depends on the number of iterations considered and the number of symbols per coherence interval. Thus, depending on these numbers, RGDPA and GDPA could have a comparable to or an even higher cost than RLSPA. Therefore, RLSPA is a low complexity technique and is attractive in scenarios where a direct and robust closed-form solution is preferred. { {In summary, RLSPA incurs a symbol-level computational load at the CPU but offers a closed-form, non-iterative and robust LS solution which it is attractive when a direct symbol-level design is desired and the number of iterations required by gradient-based block-level schemes is large. We also remark that the robust LS framework can be adapted to a block-level setting by designing a single power vector per coherence block based on symbol statistics (e.g., average power or symbol covariance) and reusing it across all symbols in the block, whose development and evaluation are left for future work.}} {In our setting, this projection-based implementation works very well and yields sum-rate performance that is very close to solutions obtained with Karush--Kuhn--Tucker (KKT) consistent non-negative least-squares (NNLS) methods, while incurring lower cost since NNLS solvers rely on iterative procedures.}
\vspace{-2mm}
\begin{algorithm}[t]
\begin{small}
\caption{ RLSPA Approach}\label{alg:RLSPA}
\begin{itemize}
  \item[] \textbf{Input: }\textup{ $\hat{\mathbf{G}}$,  $\mathbf{W}$,  $\rho_f$,  $\mathbf{x}$,} 
  \item[] \textbf{Output: }\textup{Power allocation vector $\mathbf{d}$}
\end{itemize}
\vspace{-2mm}  
\begin{enumerate}
  \item \textbf{Construct matrix $\mathbf{A}$:} \vspace{-2mm}
    \[
      \mathbf{A} \leftarrow \sqrt{\rho_f}\,\hat{\mathbf{G}}^T\,\mathbf{W}\,\mathrm{diag}(\mathbf{x}).
    \]
  \item \textbf{Compute regularization parameter $\lambda$:} \vspace{-2mm}
    \[
      \lambda \leftarrow \rho_f\,\kappa_2\,\|\mathbf{W}\,\mathrm{diag}(\mathbf{x})\|^2.
    \]
  \item \textbf{Solve for power allocation vector $\mathbf{d}$:} \vspace{-2mm}
    \[
      \mathbf{d} \leftarrow \big(\mathbf{A}^T \mathbf{A}^* + \lambda\,\mathbf{I}\big)^{-1} \mathbf{A}^T \mathbf{x}^*.
    \]
  \item \textbf{Project total power constraint:} \\
    \If{$\|\mathbf{W}\,\mathrm{diag}(\mathbf{d})\|_F^2 > P_{\mathrm{max}}$}{
      Update: \vspace{-2mm}
      \[
      \mathbf{d} \leftarrow \mathbf{d} \cdot \frac{P_{\mathrm{max}}}{\|\mathbf{W}\,\mathrm{diag}(\mathbf{d})\|_F^2}.
      \]
    }
  \item \textbf{Element-wise non-negativity:} \\
    \For{$i = 1$ to $n$}{
      Set: \vspace{-2mm}
      \[
      d_i \leftarrow \max(0, d_i).
      \]
    }
\end{enumerate}
\end{small}
\end{algorithm}

\section{Simulation Results} \label{Simul.}

\begin{figure}[t]
	\centering
		\includegraphics[width=0.975\linewidth]{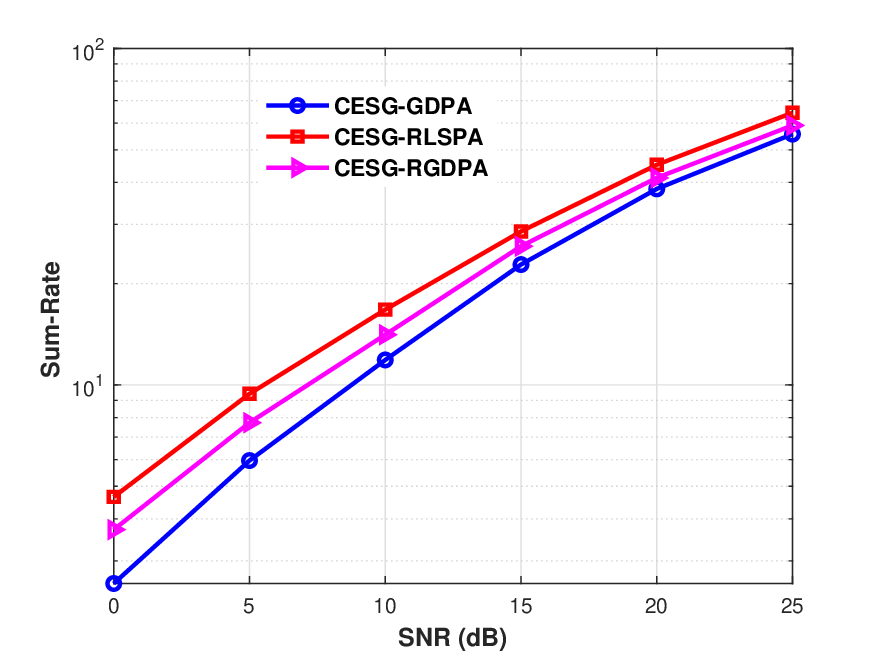}
        \vspace{-0.95em}
	\caption{\small{Sum-rate comparison of RLSPA, RGDPA, and GDPA with $\alpha=0.15$, $L=25$, $N=4$, $K=200$, $n=25$, and ZF precoding.}}
     \vspace{-0.5em}
	\label{fig:fig1}
\end{figure}

\begin{figure}[t]
	\centering
		\includegraphics[width=0.975\linewidth]{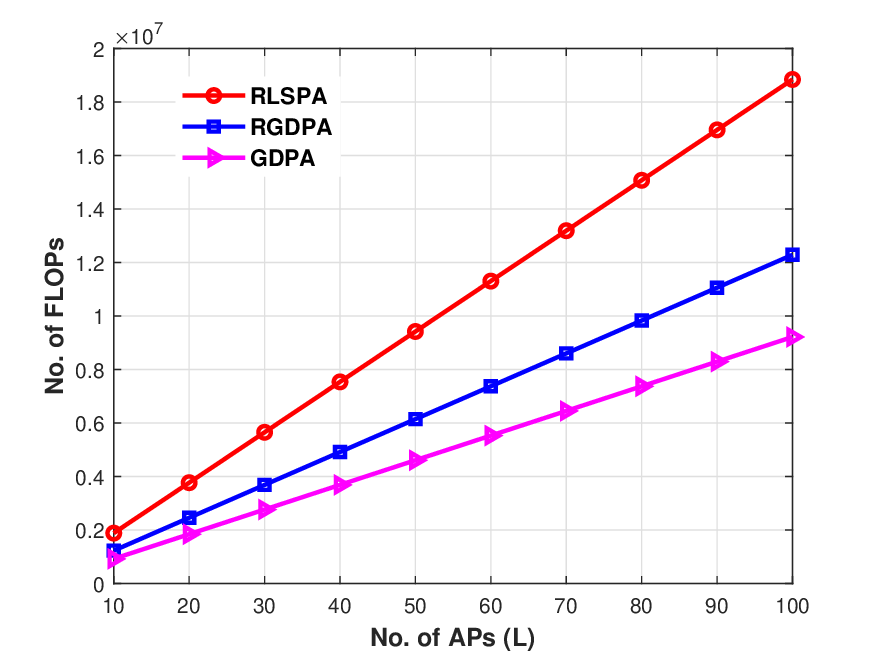}
         \vspace{-0.95em}
	\caption{\small{Complexity comparison of RLSPA, RGDPA, and GDPA in terms of FLOPs when 30 iterations are considered and $N_{\rm sym}=175$.}}
     \vspace{-0.5em}
	\label{fig:fig2}
\end{figure}

In this section, we evaluate the performance of the proposed RLSPA algorithm in the downlink of a CF-mMIMO network which is deployed in a square area of 400 meters by 400 meters, with $L = 25$ APs uniformly and independently distributed, each equipped with $N = 4$ antennas ($M = LN = 100$). $K = 200$ single-antenna UEs are randomly and uniformly distributed across the area. $n = 25$ UEs are scheduled for transmission during each coherence block. We assume that the LSFs from all antennas of an AP to a UE are identical \cite{mashdour2024clustering, nayebi2017precoding}. We assume Gaussian signaling. {All performance curves are obtained by averaging the downlink sum-rate over multiple independent Monte Carlo channel realizations.} Figure~\ref{fig:fig1} compares the sum-rate performance of RLSPA with that of the GDPA and RGDPA. All methods assume an CSI imperfection level of $\alpha = 0.15$. As SNR increases, performance of all methods is improved, however, RLSPA is superior to the other methods, confirming its effectiveness in power allocation and its robustness against CSI imperfections. RGDPA also shows better performance than GDPA as expected. Figure~\ref{fig:fig2} illustrates the computational complexity comparison between the proposed RLSPA, RGDPA, and GDPA methods, assuming 30 iterations for GDPA and RGDPA. We consider $N_{\rm sym}=175$ as the difference between the coherence interval in symbols and pilot length. As the number of APs increases, the complexity of all methods grows, with RLSPA showing slightly higher cost. In case we increase the number of iterations in GDPA and RGDPA, they can have even higher complexities than that of RLSPA. This shows that the proposed RLSPA method provides a balance between performance and computational efficiency, offering an effective and scalable solution for power allocation. 
\vspace{-2.2em}

\section{Conclusion} \label{Conclud.}

In this paper, we proposed a robust power allocation strategy for downlink CF-mMIMO systems with imperfect CSI. Using an LS formulation and incorporating Tikhonov regularization, we developed a closed-form solution that enhances robustness against channel estimation errors. The proposed RLSPA algorithm efficiently computes power allocation while satisfying power and non-negativity constraints through projection. The simulation results demonstrated that RLSPA significantly outperforms conventional methods while maintaining comparable computational complexity. This makes RLSPA a practical and scalable solution for real-world CF-mMIMO deployments.
\vspace{-1.5mm}
     \IEEEtriggeratref{1}
\IEEEtriggercmd{\vspace{-2mm}}
\bibliographystyle{IEEEbib}
\bibliography{refs}

@article{mashdour2022enhanced,
  title={Enhanced subset greedy multiuser scheduling in clustered cell-free massive MIMO systems},
  author={Mashdour, S. and de Lamare, R. C. and Lima, J. P.},
  journal={IEEE Communications Letters},
  volume={27},
  number={2},
  pages={610--614},
  year={2022},
  publisher={IEEE}
}

@article{dou2017signal,
  title={Signal restoration combining modified tikhonov regularization and preconditioning technology},
  author={Dou, H. X. and Li, H. B. and Fan, Q. Y. and Chen, Y. C.},
  journal={IEEE Access},
  volume={5},
  pages={24275--24283},
  year={2017},
  publisher={IEEE}
}

@article{fuhry2012new,
  title={A new Tikhonov regularization method},
  author={Fuhry, M. and Reichel, L.},
  journal={Numerical Algorithms},
  volume={59},
  pages={433--445},
  year={2012},
  publisher={Springer}
}

@article{2008matrix,
  title={The matrix cookbook},
  author={Petersen, K. B. and Pedersen, M. S.},
  journal={Technical University of Denmark},
  volume={7},
  number={15},
  pages={510},
  year={2008}
}

@article{hjorungnes2007complex,
  title={Complex-valued matrix differentiation: Techniques and key results},
  author={Hjorungnes, A. and Gesbert, D.},
  journal={IEEE Transactions on Signal Processing},
  volume={55},
  number={6},
  pages={2740--2746},
  year={2007},
  publisher={IEEE}
}

@article{ngo2017cell,
  title={Cell-free massive {MIMO} versus small cells},
  author={Ngo, H. Q. and Ashikhmin, A. and Yang, H. and Larsson, E. G. and Marzetta, T. L.},
  journal={IEEE transactions on wireless communications},
  volume={16},
  number={3},
  pages={1834--1850},
  year={2017},
  publisher={IEEE}
}

@inproceedings{tang2001mobile,
  title={Mobile propagation loss with a low base station antenna for NLOS street microcells in urban area},
  author={Tang, A. and Sun, J. and Gong, K.},
  booktitle={IEEE VTS 53rd Vehicular Technology Conference, Spring 2001. Proceedings (Cat. No. 01CH37202)},
  volume={1},
  pages={333--336},
  year={2001},
  organization={IEEE}
}

@inproceedings{ngo2015cell,
  title={Cell-free massive {{MIMO}}: Uniformly great service for everyone},
  author={Ngo, H. Q. and Ashikhmin, A. and Yang, H. and Larsson, E. G. and Marzetta, T. L.},
  year={2015},
  organization={IEEE 16th International Workshop on Signal Processing Advances in Wireless Communications (SPAWC)}
}

@article{interdonato2019ubiquitous,
  title={Ubiquitous cell-free massive {MIMO} communications},
  author={Interdonato, G. and Björnson, E. and Ngo, H. Q. and Frenger, P. and Larsson, E. G.},
  journal={EURASIP Journal on Wireless Communications and Networking},
  volume={2019},
  number={1},
  pages={1-13},
  year={2019},
  publisher={Springer}
}

@article{bjornson2020scalable,
  title={Scalable cell-free massive {MIMO} systems},
  author={Björnson, E. and Sanguinetti, L.},
  journal={IEEE Transactions on Communications},
  volume={68},
  number={7},
  pages={4247-4261},
  year={2020},
  publisher={IEEE}
}

@book{hofmann2013regularization,
  title={Regularization for applied inverse and ill-posed problems: a numerical approach},
  author={Hofmann, B.},
  volume={85},
  year={2013},
  publisher={Springer-Verlag}
}

@article{nayebi2017precoding,
  title={Precoding and power optimization in cell-free massive {MIMO} systems},
  author={Nayebi, E. and Ashikhmin, A. and Marzetta, T. L. and Yang, H. and Rao, B. D.},
  journal={IEEE Transactions on Wireless Communications},
  volume={16},
  number={7},
  pages={4445-4459},
  year={2017},
  publisher={IEEE}
}

@article{van2016joint,
  title={Joint power allocation and user association optimization for massive {MIMO} systems},
  author={Van Chien, T. and Björnson, E. and Larsson, E. G.},
  journal={IEEE Transactions on Wireless Communications},
  volume={15},
  number={9},
  pages={6384-6399},
  year={2016},
  publisher={IEEE}
}

@article{flores2022robust,
  title={Robust and adaptive power allocation techniques for rate splitting based {MU-MIMO} systems},
  author={Flores, A. R. and de Lamare, R. C.},
  journal={IEEE Transactions on Communications},
  volume={70},
  number={7},
  pages={4656-4670},
  year={2022},
  publisher={IEEE}
}

@article{palhares2021robust,
  title={Robust {MMSE} precoding and power allocation for cell-free massive {MIMO} systems},
  author={Palhares, V. M. T. and Flores, A. R. and de Lamare, R. C.},
  journal={IEEE Transactions on Vehicular Technology},
  volume={70},
  number={5},
  pages={5115-5120},
  year={2021},
  publisher={IEEE}
}

@article{zou2005regularization,
  title={Regularization and variable selection via the elastic net},
  author={Zou, H. and Hastie, T.},
  journal={Journal of the Royal Statistical Society Series B: Statistical Methodology},
  volume={67},
  number={2},
  pages={301-320},
  year={2005},
  publisher={Oxford University Press}
}

@inproceedings{mashdour2024robust,
  title={Robust user scheduling and power allocation in cell-free massive MIMO networks},
  author={Mashdour, S. and Flores, A. R. and Salehi, S. and de Lamare, R. C. and Schmeink, A. and Lima, J. P. and Da Silva, P. B.},
  booktitle={19th International Symposium on Wireless Communication Systems (ISWCS)},
  pages={},
  year={2024},
  organization={IEEE}
}

@article{el1997robust,
  title={Robust solutions to least-squares problems with uncertain data},
  author={El Ghaoui, L. and Lebret, H.},
  journal={SIAM Journal on matrix analysis and applications},
  volume={18},
  number={4},
  pages={1035--1064},
  year={1997},
  publisher={SIAM}
}

@article{mashdour2025robust,
  title={Robust Resource Allocation in Cell-Free Massive {MIMO} Systems},
  author={Mashdour, S. and Flores, A. R. and Salehi, S. and de Lamare, R. C. and Schmeink, A. and Da Silva, P. B.},
  journal={IEEE Transactions on Communications},
  year={2025},
  publisher={IEEE}
}

@article{flores2023clustered,
  title={Clustered Cell-Free Multi-User Multiple-Antenna Systems with Rate-Splitting: Precoder Design and Power Allocation},
  author={Flores, A. R. and de Lamare, R. C. and Mishra, K. V.},
  journal={IEEE Transactions on Communications},
  year={2023},
  publisher={IEEE}
}

@article{mishra2022rate,
  title={Rate-splitting assisted massive machine-type communications in cell-free massive {MIMO}},
  author={Mishra, A. and Mao, Y. and Sanguinetti, L. and Clerckx, B.},
  journal={IEEE Communications Letters},
  volume={26},
  number={6},
  pages={1358-1362},
  year={2022},
  publisher={IEEE}
}

@article{ammar2021downlink,
  title={Downlink resource allocation in multiuser cell-free {MIMO} networks with user-centric clustering},
  author={Ammar, H. A. and Adve, R. and Shahbazpanahi, S. and Boudreau, G. and Srinivas, K. V.},
  journal={IEEE Transactions on Wireless Communications},
  volume={21},
  number={3},
  pages={1482-1497},
  year={2022},
  publisher={IEEE}
}

@article{mashdour2024clustering,
  title={Clustering and scheduling with fairness based on information rates for cell-free {MIM} networks},
  author={Mashdour, S. and Salehi, S. and de Lamare, R. C. and Schmeink, A. and Lima, J. P.},
  journal={IEEE Wireless Communications Letters},
  year={2024},
  publisher={IEEE}
}

@ARTICLE{11311474,
  author={Mohammadzadeh, S. and M.Rahmani and Cumanan, K. and Burr, A. and Xiao, P.},
  journal={IEEE Open Journal of the Communications Society}, 
  title={Pilot and Data Power Control for Scalable Uplink Cell-Free Massive {MIMO}}, 
  year={2025},
  volume={},
  number={},
  pages={1-1},}

@inproceedings{mohammadzadeh2025association,
  title={Association of Access Points and Users and Power Allocation for Cell-Free Massive {MIMO} Systems},
  author={Mohammadzadeh, S. and Mashdour, S and de Lamare, R. C. and Cumanan, K. and Li, Ch.},
  booktitle={2025 IEEE 26th International Workshop on Signal Processing and Artificial Intelligence for Wireless Communications (SPAWC)},
  pages={1--5},
  year={2025},
  organization={IEEE}
}

@ARTICLE{decidd,
  author={Ssettumba, T. and Shao, Z. and Landau, L. T. N. and Facina, M. and da Silva, P. and de Lamare, R. C.},
  journal={IEEE Access}, 
  title={Centralized and Decentralized IDD Schemes for Cell-Free Massive MIMO Systems: AP Selection and LLR Refinement}, 
  year={2024},
  volume={12},
  number={},
  pages={62392-62406},
  doi={10.1109/ACCESS.2024.3395585}}

@ARTICLE{iddllr,
  author={Renna, R. B. Di and de Lamare, R. C.},
  journal={IEEE Transactions on Vehicular Technology}, 
  title={Iterative Detection and Decoding With Log-Likelihood Ratio Based Access Point Selection for Cell-Free MIMO Systems}, 
  year={2024},
  volume={73},
  number={5},
  pages={7418-7423},
  doi={10.1109/TVT.2023.3347097}}

@ARTICLE{iddocl,
  author={Ssettumba, T. and Mashdour, S. and Landau, L. T. N. and da Silva, P. B. and de Lamare, R. C.},
  journal={IEEE Wireless Communications Letters}, 
  title={Iterative Interference Cancellation for Clustered Cell-Free Massive MIMO Networks}, 
  year={2025},
  volume={14},
  number={2},
  pages={509-513},
  doi={10.1109/LWC.2024.3512497}}

@ARTICLE{rrs,
  author={Flores, A. R. and de Lamare, R. C.},
  journal={IEEE Communications Letters}, 
  title={Robust Rate-Splitting-Based Precoding for Cell-Free MU-MIMO Systems}, 
  year={2025},
  volume={29},
  number={6},
  pages={1230-1234},
  doi={10.1109/LCOMM.2025.3557743}}

@ARTICLE{jpba,
  author={Jiang, Y. and Zou, Y. and Guo, H. and Tsiftsis, T. A. and Bhatnagar, M. R. and de Lamare, R. C. and Yao, Y.-D.},
  journal={IEEE Transactions on Communications}, 
  title={Joint Power and Bandwidth Allocation for Energy-Efficient Heterogeneous Cellular Networks}, 
  year={2019},
  volume={67},
  number={9},
  pages={6168-6178},
  doi={10.1109/TCOMM.2019.2921022}}

@ARTICLE{rdrls,
  author={Yu, Y. and Zhao, H. and de Lamare, R. C. and Zakharov, Y. and Lu, L.},
  journal={IEEE Transactions on Signal Processing}, 
  title={Robust Distributed Diffusion Recursive Least Squares Algorithms With Side Information for Adaptive Networks}, 
  year={2019},
  volume={67},
  number={6},
  pages={1566-1581},
  doi={10.1109/TSP.2019.2893846}}

@ARTICLE{tds,
  author={Clarke, P. and de Lamare, R. C.},
  journal={IEEE Communications Letters}, 
  title={Joint Transmit Diversity Optimization and Relay Selection for Multi-Relay Cooperative MIMO Systems Using Discrete Stochastic Algorithms}, 
  year={2011},
  volume={15},
  number={10},
  pages={1035-1037},
  doi={10.1109/LCOMM.2011.082611.102262}}

@ARTICLE{rthp,
  author={Zhang, L. and Cai, Y. and de Lamare, R. C. and Zhao, M.},
  journal={IEEE Transactions on Communications}, 
  title={Robust Multibranch Tomlinson–Harashima Precoding Design in Amplify-and-Forward MIMO Relay Systems}, 
  year={2014},
  volume={62},
  number={10},
  pages={3476-3490},
  doi={10.1109/TCOMM.2014.2359438}}

@ARTICLE{lrcc,
  author={Ruan, H. and de Lamare, R. C.},
  journal={IEEE Transactions on Signal Processing}, 
  title={Distributed Robust Beamforming Based on Low-Rank and Cross-Correlation Techniques: Design and Analysis}, 
  year={2019},
  volume={67},
  number={24},
  pages={6411-6423},
  doi={10.1109/TSP.2019.2954519}}

@ARTICLE{rsthp,
  author={Flores, A. R. and De Lamare, R. C. and Clerckx, B.},
  journal={IEEE Transactions on Communications}, 
  title={Tomlinson-Harashima Precoded Rate-Splitting With Stream Combiners for MU-MIMO Systems}, 
  year={2021},
  volume={69},
  number={6},
  pages={3833-3845},
  doi={10.1109/TCOMM.2021.3065145}}

@ARTICLE{mmimo,
  author={de Lamare, R. C.},
  journal={URSI Radio Science Bulletin}, 
  title={Massive MIMO systems: Signal processing challenges and future trends}, 
  year={2013},
  volume={2013},
  number={347},
  pages={8-20},
  doi={10.23919/URSIRSB.2013.7909827}}

@ARTICLE{wence,
  author={Zhang, W. and Ren, H. and Pan, C. and Chen, M. and de Lamare, R. C. and Du, B. and Dai, J.},
  journal={IEEE Transactions on Communications}, 
  title={Large-Scale Antenna Systems With UL/DL Hardware Mismatch: Achievable Rates Analysis and Calibration}, 
  year={2015},
  volume={63},
  number={4},
  pages={1216-1229},
  doi={10.1109/TCOMM.2015.2395432}}

@ARTICLE{locsme,
  author={Ruan, H. and de Lamare, R. C.},
  journal={IEEE Signal Processing Letters}, 
  title={Robust Adaptive Beamforming Using a Low-Complexity Shrinkage-Based Mismatch Estimation Algorithm}, 
  year={2014},
  volume={21},
  number={1},
  pages={60-64},
  doi={10.1109/LSP.2013.2290948}}

@ARTICLE{tds2,
  author={Clarke, P. and de Lamare, R. C.},
  journal={IEEE Transactions on Vehicular Technology}, 
  title={Transmit Diversity and Relay Selection Algorithms for Multirelay Cooperative MIMO Systems}, 
  year={2012},
  volume={61},
  number={3},
  pages={1084-1098},
  doi={10.1109/TVT.2012.2186619}}

@ARTICLE{okspme,
  author={Ruan, H. and de Lamare, R. C.},
  journal={IEEE Transactions on Signal Processing}, 
  title={Robust Adaptive Beamforming Based on Low-Rank and Cross-Correlation Techniques}, 
  year={2016},
  volume={64},
  number={15},
  pages={3919-3932},
  doi={10.1109/TSP.2016.2550006}}

@ARTICLE{rsbd,
  author={Flores, A. R. and de Lamare, R. C. and Clerckx, B.},
  journal={IEEE Communications Letters}, 
  title={Linear Precoding and Stream Combining for Rate Splitting in Multiuser MIMO Systems}, 
  year={2020},
  volume={24},
  number={4},
  pages={890-894},
  doi={10.1109/LCOMM.2020.2969158}}
\end{document}